%% file: BinaryGFH-v2.tex
\documentclass[aps,prd,twocolumn,nofootinbib,superscriptaddress]{revtex4-1}
\usepackage[utf8]{inputenc}

\usepackage{color}
\usepackage{graphicx}
\usepackage{amsmath}
\usepackage{amssymb}
\usepackage{bm}
\usepackage{acronym}
\usepackage{ifthen}
\usepackage{blindtext}
\usepackage[normalem]{ulem}
\usepackage{hyperref}
\usepackage[capitalise]{cleveref}
\usepackage{etoolbox}
\usepackage{xspace}
\usepackage{textcomp}
\usepackage{multirow}
\usepackage{lineno}
\usepackage{tabularx}
\usepackage{subfigure}
\usepackage{orcidlink}

\def\gw{GW\xspace}
\def\gwh{GW\xspace}
\def\cwh{CW\xspace}

\def\tcw{tCW\xspace}
\def\tcws{tCWs\xspace}
\def\cws{CWs\xspace}
\def\gws{GWs\xspace}

\def\nss{neutron stars\xspace}

\def\bhs{black holes\xspace}

\def\ce{Cosmic Explorer\xspace}
\def\dm{DM\xspace}
\def\dmh{DM\xspace}
\def\pbh{PBH\xspace}
\def\pbhs{PBHs\xspace}

\def\bnsh{binary neutron-star\xspace}
\def\fh{\emph{frequency-Hough}\xspace}
\def\lvk{LIGO, Virgo and KAGRA\xspace}
\def\ifo{interferometer\xspace}
\def\ifos{interferometers\xspace}

\def\invkpccubedyr{kpc$^{-3}$yr$^{-1}$\xspace}

\def\mf{matched filtering\xspace}
\def\mfh{matched-filtering\xspace}
\def\psd{power spectral density\xspace}
\def\psds{power spectral density\xspace}
\def\nn{\nonumber}

\def\gfh{\emph{GFH}\xspace}
\def\GFHvone{\emph{GFH}-v1\xspace}
\def\GFHvtwo{\emph{GFH}-v2\xspace}
\def\BinGFH{\texttt{BinaryGFH}-v2\xspace}

\def\OFourA{O4a\xspace}

\def\snr{signal-to-noise ratio\xspace}

\def\pmap{\emph{peakmap}\xspace}
\def\pmaps{\emph{peakmaps}\xspace}
\def\hm{\emph{Hough map}\xspace}
\def\hms{\emph{Hough maps}\xspace}
\def\erfc{\mathrm{erfc}}
\def\homin{h^{\Gamma}_{0,\rm min}}
\def\dmax{d^\Gamma_{\rm max}}
\def\ncs{number counts\xspace}

\newcommand{\bea}{\begin{eqnarray}}
\newcommand{\eea}{\end{eqnarray}}
\newcommand{\be}{\begin{equation}}
\newcommand{\ee}{\end{equation}}
\newcommand{\thetathr}{\theta_\text{thr}}

\newcommand{\hz}{\,\mathrm{Hz}\,}

\newcommand{\avgVT}{\ensuremath{\left\langle VT \right\rangle}}

\newcommand{\TFFT}{T_\text{FFT}}
\newcommand{\fgw}{f_\text{GW}}

\newcommand{\fdot}{\dot{f}}
\newcommand{\Tobs}{T_\text{obs}}
\newcommand{\Tpm}{T_\text{PM}}
\newcommand{\fpbh}{f_\text{PBH}}

\newcommand{\ftilde}{\tilde{f}}

\newcommand{\fsup}{f_\text{sup}}
\newcommand{\fmone}{f(\ln m_1)}
\newcommand{\fmtwo}{f(\ln m_2)}
\newcommand{\Mc}{\mathcal{M}}

\newcommand{\fmin}{f_\text{min}}
\newcommand{\fmax}{f_\text{max}}
\newcommand{\ssm}{sub-solar-mass\xspace}
\newcommand{\msun}{\ensuremath{M_\odot}\xspace}
\newcommand{\crthr}{CR_\text{thr}}

\newcommand{\mpbh}{M_\text{PBH}}

\newcommand{\gnlvl}{S_n(f)=7.94\times 10^{-24}\, \hz^{-1/2}}

\newtoggle{fullauthorlist}
\toggletrue{fullauthorlist}
\newtoggle{endauthorlist}
\toggletrue{endauthorlist}

\newcommand{\tcr}{}
\newcommand{\tcb}{}

\graphicspath{{./figures/}}

\begin{document}
\title{\texttt{BinaryGFH-v2}: Improved method to search for gravitational waves from sub-solar-mass, ultra-compact binaries using the Generalized Frequency-Hough Transform}
\author{Andrew L. Miller\,\orcidlink{0000-0002-4890-7627}}
\email{andrew.miller.ligo@ucas.ac.cn}
\affiliation{International Centre for Theoretical Physics Asia-Pacific (ICTP-AP), University of Chinese Academy of Sciences (UCAS), Beijing 100190, China.}
\affiliation{Taiji Laboratory for Gravitational Wave Universe, University of Chinese Academy of Sciences, 100049 Beijing, China}
\author{Lorenzo Pierini}
\affiliation{INFN, Sezione di Roma, I-00185 Roma, Italy}
\date{\today}

\begin{abstract}

Observing gravitational waves from sub-solar-mass, inspiraling compact binaries would provide almost smoking-gun evidence for primordial black holes. Here, we develop a method to search for ultra-compact binaries with chirp masses ranging from $\mathcal{M}\in [10^{-2},10^{-1}]M_\odot$. This mass range represents a previously unexplored gap in gravitational-wave searches for compact binaries: it was thought that the signals would too long for matched-filtering analyses but too short for time-frequency pattern-recognition techniques. Despite this, we show that a pattern-recognition technique, the \emph{Generalized frequency-Hough} (GFH), can be employed with particular modifications that allow us to handle rapidly spinning-up binaries and to increase the statistical robustness of our method, and call this improved method \texttt{BinaryGFH-v2}. We then design a hypothetical search for binaries in this mass regime, compare the empirical and theoretical sensitivities of this method, and project constraints on formation rate densities and the fraction of dark matter that primordial black holes could compose in both current- and future-generation gravitational-wave detectors. Our results show that our method can be used to search for sub-solar-mass ultra-compact objects in a mass regime that remains to-date unconstrained with gravitational waves.
\end{abstract}
        
        \maketitle

\section{Introduction}\label{sec:intro}

Despite indirect evidence for the existence of dark matter (\dm), decades of research have been unable to elucidate its true nature. Potential masses for the constituent \dmh particles range from $\mathcal{O}(10^{-22})$ eV to $\mathcal{O}(100)$ GeV, which have been proposed to be axions \cite{Abbott:1982af,Preskill:1982cy}, Weakly Interacting Massive Particles (WIMPs) \cite{Jungman:1995df}, and sterile neutrinos \cite{Boyarsky:2018tvu}, among many others. In contrast, macroscopic \dm, e.g. in the form of primordial black holes (\pbhs) that could have formed in the early universe through the collapse of overdensities \cite{Hawking:1971ei,Khlopov:2008qy,Carr:2016drx,Carr:2019kxo, Escriva:2022duf}, could be as light as $\mathcal{O}(10^{-18})\msun$ and as heavy as $\mathcal{O}(10^3)\msun$. Moreover, the 218 detections of \gws from merging \bhs \cite{LIGOScientific:2025slb}, whose rates and spins are consistent with those expected from \pbhs \cite{Clesse:2017bsw,Clesse:2020ghq}, have revived both theoretical and experimental interest in \pbhs. Taken together, the possible mass of \dm covers approximately ninety orders of magnitude.

Recently, it has been shown that gravitational-wave (\gwh) \ifos can probe the existence of \dm -- see \cite{Bertone:2019irm,Green:2020jor,Bertone:2024rxe,Miller:2024rca,Miller:2025yyx} for topical reviews. In particular, searches for \gws from ultra-compact objects in binaries with masses $m_1,m_2$, the so-called \ssm regime ($m_1 \ \text{or}\ m_2<\msun$), have been performed to probe the existence of \pbhs and dark \bhs \cite{LIGOScientific:2018glc,LIGOScientific:2019kan,Singh:2020wiq,Nitz:2020bdb,LIGOScientific:2021job,Phukon:2021cus,Nitz:2021mzz,Nitz:2021vqh,LVK:2022ydq,LIGOScientific:2022hai,Nitz:2022ltl,Yuan:2024yyo,Miller:2024fpo,LIGOScientific:2025vwc}. In this mass regime, no astrophysical mechanisms exist to form such \bhs;  thus, detecting the inspiral or merger of such systems would provide almost unambiguous evidence for a primordial origin of such compact objects \cite{LISACosmologyWorkingGroup:2023njw,Yamamoto:2023tsr}.

Searches for \ssm, ultra-compact binaries with chirp masses $\Mc=[10^{-1},1]\msun$ largely rely on \mf, a technique that correlates a large template bank of possible signals with the data and outputs a detection statistic, called the \snr, that quantifies how well that template matches the data. This technique is ideal if one completely trusts their model and has large amounts of computing power to spare: in practice, \mfh searches correlate millions of templates, of durations of $\mathcal{O}(100)$ s or less, over the year-long observation times $\Tobs$ of the detector. The vast parameter space covered in \mfh searches, coupled with the long observing runs of \lvk \cite{2015CQGra..32g4001L,2015CQGra..32b4001A,KAGRA:2020tym}, imply a massive computational cost, which only increases as the chirp mass $\Mc$ of the system decreases. As $\Mc$ decreases, the signal spends more time in the frequency band of the detector, and thus the accumulated phase mismatches over time between two templates close to each other in the parameter space also increase. Therefore, more templates are needed the parameter space to guarantee a chosen mismatch between templates.

The excessive computational cost has limited all \mfh method to search for systems with $\Mc\gtrsim10^{-1}\msun$\footnote{We note that one search was performed in O1 and O2 for compact objects with a \emph{secondary} masses as low as $10^{-2}\msun$ \cite{Nitz:2020bdb} in the specific case of highly asymmetric mass-ratio systems, though the minimum searched chirp mass was $10^{-1}\msun$.}. Recently, however, it was determined in \cite{Miller:2020kmv} that \ssm binaries in the mass range $\Mc=[10^{-5},10^{-2}]\msun$ would emit \gws during their inspiral for $\mathcal{O}(\rm hours-days)$, and can thus be probed with pattern-recognition techniques that attempt to find tracks in time--frequency representations of the data, where each track corresponds to a unique chirp mass. The method in \cite{Miller:2020kmv} derives from so-called transient continuous-wave (\tcw) methods \cite{Prix:2011qv,Oliver:2018dpt,Sun:2018owi,Banagiri:2019obu,keitel2019first} that were originally designed to search for remnants of \bnsh mergers \cite{longpmr}, which are themselves derivatives of methods to search for isolated, asymmetrically rotating \nss \cite{Jaranowski:1998qm,Krishnan:2004sv,Sintes:2006uc,Astone:2014esa,Keitel:2015ova,Suvorova:2016rdc,LIGOScientific:2019yhl}. Other efforts followed to search for compact objects inspiraling in this mass regime \cite{VelcaniThesis,Andres-Carcasona:2023zny,Andres-Carcasona:2024jvz,Alestas:2024ubs}. 

Moreover, binaries with $\Mc=[10^{-7},10^{-5}]\msun$ can inspiral for years, if not longer, and can thus be probed with traditional continuous-wave (\cwh) methods in all-sky searches for these binaries \cite{Miller:2021knj,KAGRA:2022dwb}. These methods are orders of magnitude more computationally efficient than \mf, but incur a small sensitivity loss around a factor of a few depending on the specific choices made in an analysis and the available computing power budget \cite{Astone:2014esa}. However, this loss is necessary if we wish to be able to perform searches for \ssm objects with $\Mc\ll10^{-1}\msun$.

From the above discussion, we see a clear gap at chirp masses of $\Mc=[10^{-2},10^{-1}]\msun$. Thus, our aim is to develop a technique that can be used to probe this mass regime, which remains unexplored by both current \tcw and \mfh searches. These signals were thought to be ``too short'' for \tcw searches but ``too long'' for \mfh searches. Here, we show that, with specific modifications to a particular \tcw search pipeline, the \emph{Generalized frequency-Hough} (GFH) \cite{Miller:2020kmv,Miller:2024jpo}, we can access the full chirp mass range of $\Mc=[10^{-2},10^{-1}]\msun$.

In the context of \pbhs, we note that stringent constraints already exist on the fraction of \dm that \pbhs could compose in this mass range \cite{Green:2020jor}. These constraints come from gravitational lensing of the light from distant galaxies or supernovas induced by hypothetical \pbhs that sit between us and the source \cite{EROS-2:2006ryy,Niikura:2019kqi,Croon:2020ouk}. However, these constraints are model dependent: they assume monochromatic mass functions and can be weakened if \pbhs form in clusters. In particular, isolated \pbhs would be less likely to fall along the line-of-sight between earth and the Large Magellenic Cloud compared to uniformly distributed \pbhs, thus reducing the probability of microlensing events \cite{Clesse:2016vqa,Garcia-Bellido:2017xvr}. Thus, using different experimental probes, and assuming different formation mechanisms for \pbhs (probing binaries with \gws versus isolated ones with microlensing) can provide complementary, robust constraints on the existence of \pbhs across these mass regimes.

We divide this paper into the following sections: in \cref{sec:sig}, we describe the properties of the \gwh signal from inspiraling compact object that we consider throughout this work. In \cref{sec:gfh}, we explain the original method, the \GFHvone, the second version of it, \GFHvtwo, and its improved version, the \BinGFH, to handle signals with  $\Mc=[10^{-2},10^{-1}]\msun$. Then, in \cref{sec:search-design}, we propose a possible search design using the \BinGFH for an analysis of O4a data. Afterwards, in \cref{sec:sens}, we evaluate the sensitivity of our new method with injections and compare it to \GFHvone, \GFHvtwo and the theoretical sensitivity; in \cref{sec:proj-constr}, we calculate the expected sensitivity we will have in both O4a and \ce \cite{Reitze:2019iox}, and show what kinds of constraints we can obtain on the fraction of \dm that \pbhs can compose. Finally, in \cref{sec:concl}, we make some concluding remarks and propose avenues for future work.

\section{The Signal}\label{sec:sig}

Two compact objects in orbit around their center of mass will emit \gws as they approach each other. Equating the orbital energy loss with \gwh power, we can obtain the rate of change of the frequency over time, i.e. the spin-up $\fdot$, in the quasi-Newtonian limit (i.e. far from merger) \cite{maggiore2008gravitational}:

\begin{align}
    \dot{f}_{\rm GW}&=\frac{96}{5}\pi^{8/3}\left(\frac{G\Mc}{c^3}\right)^{5/3} \fgw^{11/3}\equiv k \fgw^{11/3} \nonumber \\
    &\simeq 5.8\times 10^{-3}\text{ Hz/s} \left(\frac{\Mc}{10^{-2}M_\odot}\right)^{5/3}\left(\frac{\fgw}{100\text{ Hz}}\right)^{11/3},
    \label{eqn:fdot_chirp}
\end{align}
where $\Mc\equiv\frac{(m_1m_2)^{3/5}}{(m_1+m_2)^{1/5}}$ is the chirp mass of the system, $\fgw$ is the \gwh frequency, $c$ is the speed of light, and $G$ is Newton's gravitational constant.

To obtain the signal frequency evolution $\fgw(t)$ over time, we can integrate \cref{eqn:fdot_chirp} with respect to time $t$:
\begin{equation}
\fgw(t)=f_0\left[1-\frac{8}{3}kf_0^{8/3}(t-t_0)\right]^{-\frac{3}{8}}~,
\label{eqn:powlaws}
\end{equation}
where $t_0$ is a reference time for the \gwh frequency $f_0$. 

The amplitude $h_0(t)$ of the \gwh signal also evolves with time \cite{maggiore2008gravitational}:

\begin{align}
h_0(t)&=\frac{4}{d}\left(\frac{G \Mc}{c^2}\right)^{5/3}\left(\frac{\pi \fgw(t)}{c}\right)^{2/3} \nonumber \\
&\simeq 1.2\times 10^{-24}\left(\frac{1\text{ Mpc}}{d}\right)\left(\frac{\Mc}{10^{-2}\msun}\right)^{5/3}\left(\frac{\fgw}{100\text{ Hz}}\right)^{2/3},
\label{eqn:h0}
\end{align}
where $d$ is the luminosity distance to the source.

\tcb{The signal $h(t)$ received at the detector is obtained by calculating how the \ifos responds to the incoming \gw:
\begin{equation}
    h(t)=F_+(t,\alpha,\beta,\psi)h_+(t,\iota) \, + F_\times(t,\alpha,\beta,\psi) h_\times(t,\iota)
    \label{eqn:antenna}
\end{equation}
where $F_+$ and $F_\times$ are the beam pattern functions of the detector and are given in \cite{Jaranowski:1998qm}, and $h_+$ and $h_\times$ are the two polarizations of the \gwh signal, and can be written as:
\begin{align}
    h_+ &=h_0\left(\frac{1+\cos^2\iota}{2} \right)\\
    h_\times &= h_0\cos^2\iota
\end{align}}

Inverting \cref{eqn:powlaws}, we can also write down an expression for the time the signal spends between two frequencies:

\begin{equation}
    t-t_0\equiv\Delta t=-\frac{3}{8}\frac{\fgw^{-8/3}-f_0^{-8/3 }}{k},
\end{equation}
which, in the limit that $\fgw\rightarrow\infty$, determines the time to merger $t_{\rm coal}$

\begin{align}
    t_{\rm coal}&\simeq\frac{5}{256}\left(\frac{1}{\pi f_0}\right)^{8/3}\left(\frac{c^3}{G\Mc}\right)^{5/3}\nonumber \\
    &\simeq 6470\text{ s} \left(\frac{100\text{ Hz}}{f_0}\right)^{8/3} \left(\frac{10^{-2}\msun}{\Mc}\right)^{5/3}.
    \label{eqn:tmerg}
\end{align}
As seen above, systems with $\Mc\ll\msun$ spend a long time in the detector frequency band before merging relative to those with $\Mc\gtrsim\msun$. 
Such long-duration signals are problematic for conventional matched-filtering algorithms, thus motivating the development of alternative methods to search for such light-mass \pbhs \cite{Miller:2020kmv,Andrés-Carcasona:2023df,Alestas:2024ubs}.

In \cref{fig:mergtime}, we plot time to merger as a function of chirp mass (\cref{eqn:tmerg}). While \mf can look for signals between $[0.1,1]\msun$ \cite{LIGOScientific:2022hai}, \tcw methods, such as the \gfh \cite{Miller:2018rbg,Miller:2020kmv} or other pattern-recognition techniques \cite{Sun:2018hmm,Oliver:2018dpt,Miller:2019jtp,Andres-Carcasona:2024jvz,Andres-Carcasona:2024jvz,Alestas:2024ubs}, can probe $[10^{-5},10^{-1}]\msun$, while traditional \cwh methods, such as the original \fh \cite{Astone:2014esa}, can probe $[10^{-7},10^{-5}]\msun$. 

\begin{figure}
    \centering
\includegraphics[width=\columnwidth]{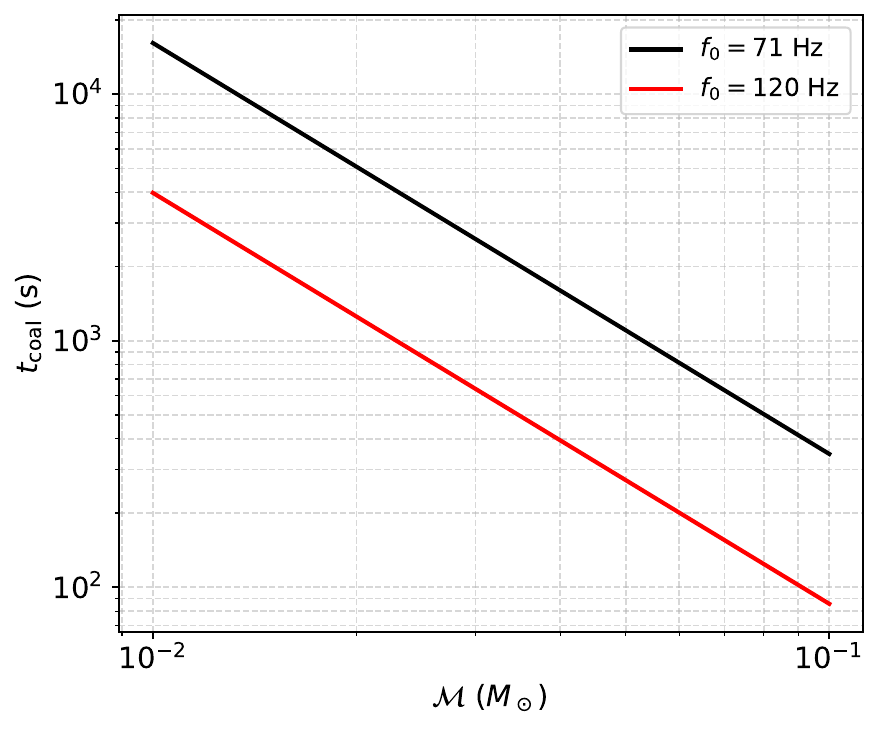}
    \caption{\textbf{Time to merger as a function of chirp mass for initial frequencies considered in this paper.} }
    \label{fig:mergtime}
\end{figure}

\section{Generalized Frequency-Hough Transform}\label{sec:gfh}

\subsection{Creation of the \pmap}

Typically, \tcw methods take as input time--frequency representations of the data and output a statistic that indicates to what extent an astrophysical signal is present in the data. In this case, we apply our method, the \gfh, on a time--frequency \pmap, which is created by breaking the data, of duration $\Tpm$, into shorter durations of length $\TFFT$, fast Fourier transforming them, estimating the \psd in each of these chunks, and dividing the FFT power by the \psd as a way of equalizing or whitening the data. On this equalized power, which has a mean and standard deviation equal to one, we require that pixels in time--frequency pass a threshold that indicates the number of standard deviations away from the mean, typically $\thetathr=2.5$, and that they are local maxima. The output of this procedure is called a \pmap, which is a time--frequency map containing only points (``peaks'') that pass the aforementioned criteria. In our case, we build these \pmaps from short fast Fourier Transform databases (SFDBs) \cite{Astone:2005fj}, which contain FFTs of 1024 s. Practically, we inverse Fourier transform these 1024-second FFTs to the time-domain and then re-FFT them with our desired $\TFFT$. More details on the construction of the \pmap can be found in \cite{Astone:2005fj,Astone:2014esa,Miller:2018rbg,Miller:2020kmv,Miller:2024jpo}.

\tcb{At this point, analyses for \cws typically weight the \pmap by the particular antenna response ($F_+$ and $F_\times$ referenced in \cref{eqn:antenna}) based on when the data were taken and what sky position the signal is assumed to come from. However, because the signal duration and analysis coherence time in this paper are both so short with respect to those studied in standard \cwh analyses, 
\tcr{\BinGFH does not explicitly search over sky position and does not apply antenna-pattern weighting or Doppler corrections for particular source locations. The search is therefore ``all-sky'' in the sense that no template bank over sky position is used; however, the antenna-pattern responses naturally induce a sky-dependent sensitivity, and some locations will have reduced sensitivity compared to others.}

When we compute sensitivity estimates in Gaussian noise, as done in \cref{sec:sens}, we fix the absolute times at which we perform the injections. In contrast, if we were to perform a search in a given observing run, we would obtain robust upper limits by injecting signals at different times so that our sensitivity properly account for variations in the detector response.}

\subsection{\GFHvone: Original method}

The \GFHvone \cite{Miller:2018rbg,Miller:2024jpo} operates on a time--frequency representation of \ifo data and tracks signals whose time--frequency evolution follows a power law. In particular, it works by transforming the frequencies of the \pmap as:

\begin{equation}
x=\fgw^{1-n} ;\,\,\,\, x_0=f_0^{1-n}.\label{eqn:xx0}
\end{equation}
At this point, the plane of the \pmap is no longer $t-f$ but $t-x$, and \cref{eqn:powlaws} can be written as

\begin{equation}
x=x_0-\frac{8}{3}k(t-t_0),
\end{equation}
where we have written $n=11/3$ for the case of \gwh emissions from inspiraling ultra-compact binaries.

Different tracks in the \pmap correspond to different $f_0-\Mc$ pairs, and the number of peaks that fall along each track is histogrammed in the \hm.

In a real search, the \pmap is composed of times and frequencies separated by $\TFFT/2$ and $1/\TFFT$, respectively, and thus the \hm is discretized in a specific way. Conceptually, we have to decide how many time--frequency tracks we wish to search over, which corresponds to the number of $x_0$ and $k$ values that define the axes of the \hm. In \cite{Miller:2018rbg,Miller:2024jpo}, we determine the grid in $x_0$ as:

\begin{equation}
    dx_0 = (n - 1) \frac{\delta f}{\fgw},
    \label{eqn:dx0}
\end{equation}
where $dx_0$ is the step in $x_0$, and we set $\fgw=\fmax$ in order to have a uniform, over-resolved grid. We then calculate the grid in $k$ as:
\begin{equation}
    dk = k \left( \left( 1 + \frac{\delta f}{\fmax} \right)^{-n} - 1 \right),
    \label{eqn:dk}
\end{equation}
where $dk$ is the step in $k$ that is also a function of $k$ and thus is not uniform.

Each choice for $x_0$ and $k$ uniquely determines a line in the \pmap, over which the number of peaks that fall along that line are summed and histogrammed in the \hm. 

From the \hm, we calculate a detection statistic, called the critical ratio $CR$:

\begin{equation}
    CR=\frac{n-\mu}{\sigma}
\end{equation}
where $n$ is the number of peaks that accumulate in a given pixel, $\mu$ and $\sigma$ are the mean and standard deviation of the number of peaks of the map (or a local region around which $n$ is determined).

\subsection{\GFHvtwo: Order-of-magnitude speed-up}

Recently, the \GFHvone was modified to lessen its computational cost by an order of magnitude \cite{Menon:2025wce}. In particular, the main advancement is being able to calculate the number of peaks at all $x_0$ over all times $t$ at each slope ($k$) of the \hm in parallel. This amounts to vectorizing the previous double loop over $t$ and $k$ present in the \GFHvone without a significant increase in allocated memory. We describe below the salient reasons for this speed-up.

In the standard \GFHvone implementation, each time segment of the \pmap is iterated sequentially, with its peaks contributing to a common \hm. Since each time column writes to overlapping memory locations, this formulation could be efficiently parallelized, unless one wishes to use significant amounts of memory.

The accelerated implementation achieves an order-of-magnitude speed-up by inverting the loop structure of the algorithm. Instead of iterating over times, the trick is to iterate over \hm columns corresponding to different $k$ values. For each column, the entire \pmap is shifted according to the expected frequency evolution at that $k$ over all times and $x_0$, and the result is accumulated into a one-dimensional histogram representing a vertical slice of the \hm. This ``loop inversion'' makes the computation for each column independent, allowing full parallelization over the grid of $k$.

To further optimize performance, all coordinates are pre-normalized to integer grid indices, and accumulation is performed using the vectorized MATLAB function \texttt{accumarray}. This eliminates explicit nested loops and minimizes floating-point operations, reducing both runtime and memory usage. In practice, the combination of loop inversion, coordinate normalization, and vectorized accumulation reduces the computational cost by roughly an order of magnitude with respect to the \GFHvone, while producing an identical \hm to that from \GFHvone.

\subsection{\BinGFH: Improved method developed in this work}

We aim to alleviate two limitations of the \GFHvone and \GFHvtwo. The first is that the distribution of the number counts in the \hm does not obey a normal distribution, in contrast to the original \fh \cite{Astone:2014esa}. In \cref{fig:hm-gauss-noise-uni-problem}, we show that, even for a \pmap created in Gaussian noise, the distribution of the \ncs in the \hm is bimodal. This is a problem if we wish to infer the meaning of any statistic derived from the \hm. While the \GFHvone and \GFHvtwo can successfully detect simulated signals and can be applied in real searches, we have not yet entered the regime in which we would have to confirm a detection with a given statistical confidence level.

The second limitation is that signal power is spread across the over-resolved $x_0$ grid in the Hough plane. In \cref{fig:hm-uni-inj-gauss-problem}, we can observe this reality in the \hm: though the greatest number of peaks that can accumulate in a pixel in the \hm is $N_{\rm max}=2(\frac{\Tpm}{\TFFT})=4633$, fewer than 100 peaks accumulate in the pixels surrounding the injection parameters. Of course, noise power is also spread across many pixels, but this over-resolution of the grid in $x_0$ has particular consequences for analyses in very low frequency bands, e.g. $[2,10]$ Hz) and analyses for very rapidly chirping signals, as will be shown in \cref{sec:sens}.

\begin{figure*}[ht!]
     \begin{center}
        \subfigure[ ]{%
            \label{fig:hm-uni-hough-gauss}
            \includegraphics[width=0.49\textwidth]{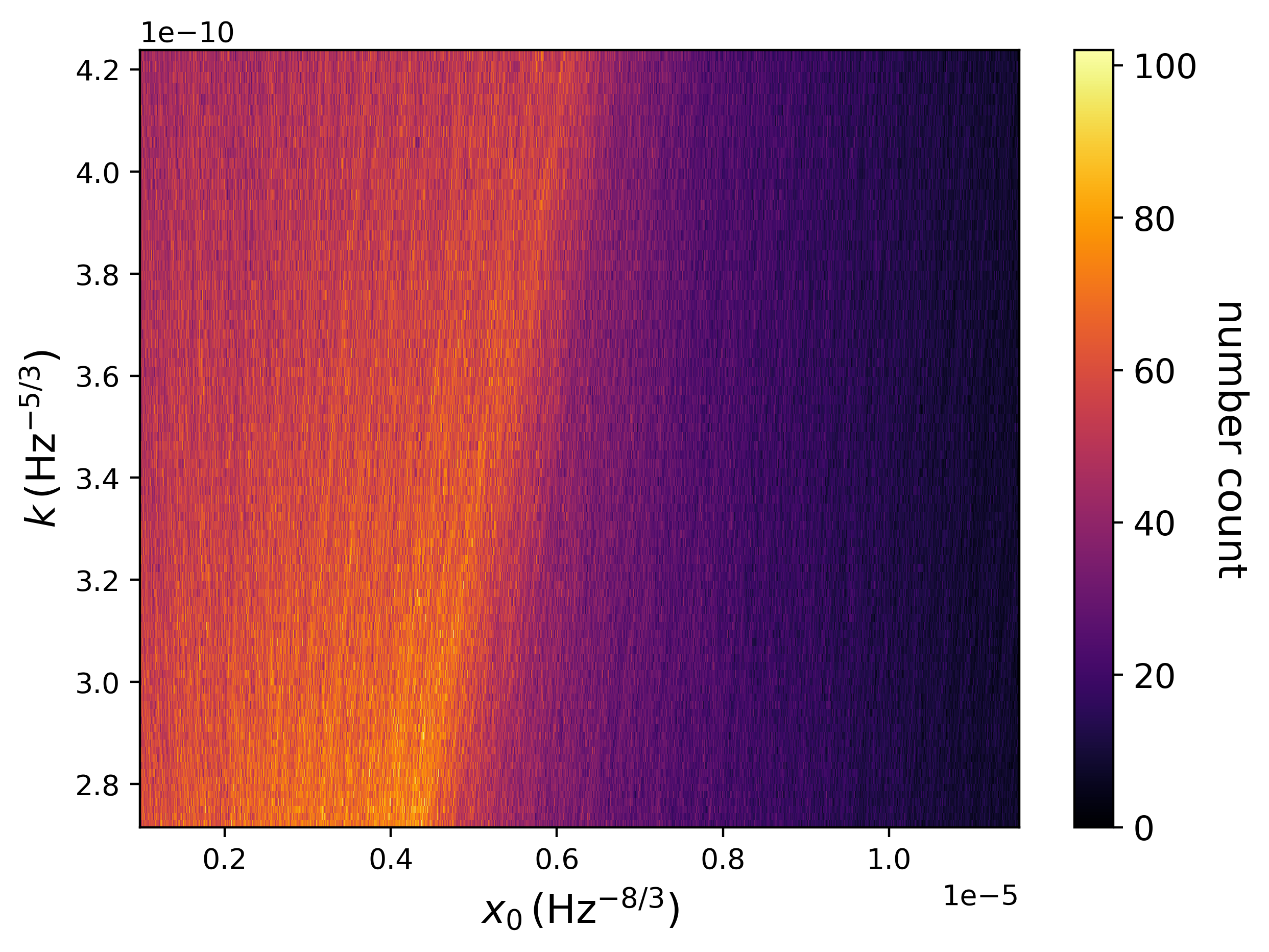}
        }%
        \subfigure[]{%
           \label{fig:nc-dist-uni-hough}
           \includegraphics[width=0.5\textwidth]{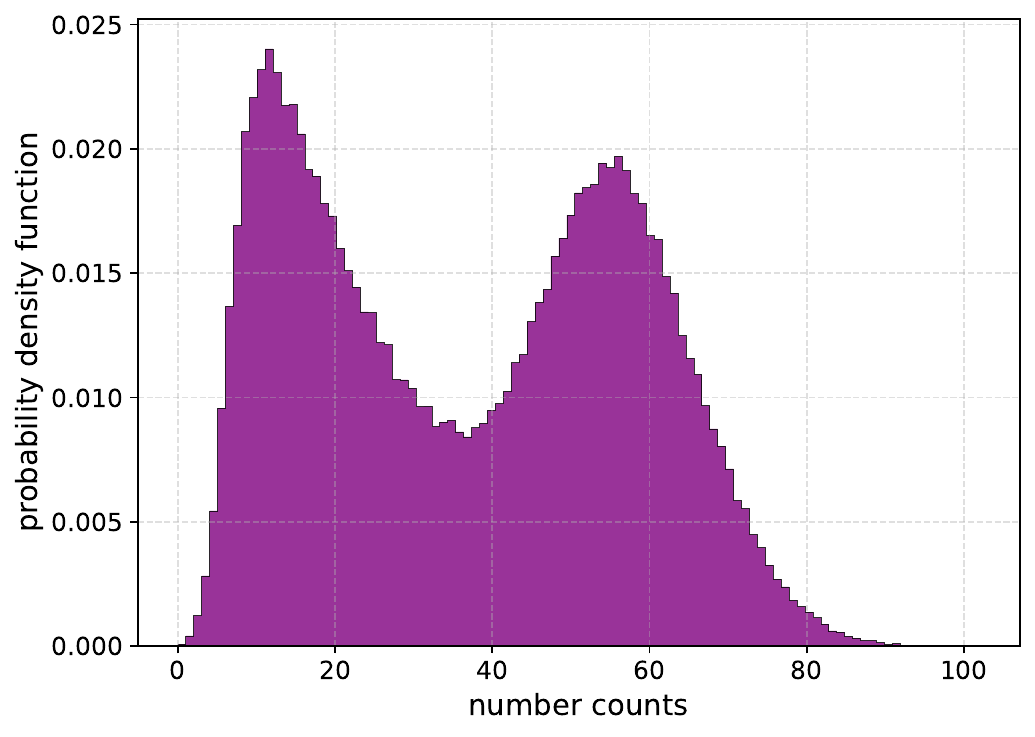}
        }\\ 
    \end{center}
    \caption[]{\textbf{Limitation 1: Non-Gaussian distribution of number counts in the \hm even in pure Gaussian noise.} Left: \hm in Gaussian noise created with a uniform grid in $x_0$ using \GFHvtwo. Right: histogram of number counts. $\TFFT=4$ s; $\Tpm=9266$ s; $\Mc=[1,1.3]\times 10^{-2}\msun$; $\fgw=[71,169]$ Hz. 
    }%
   \label{fig:hm-gauss-noise-uni-problem}
\end{figure*}

\begin{figure}[ht!]
     \begin{center}
\includegraphics[width=0.49\textwidth]{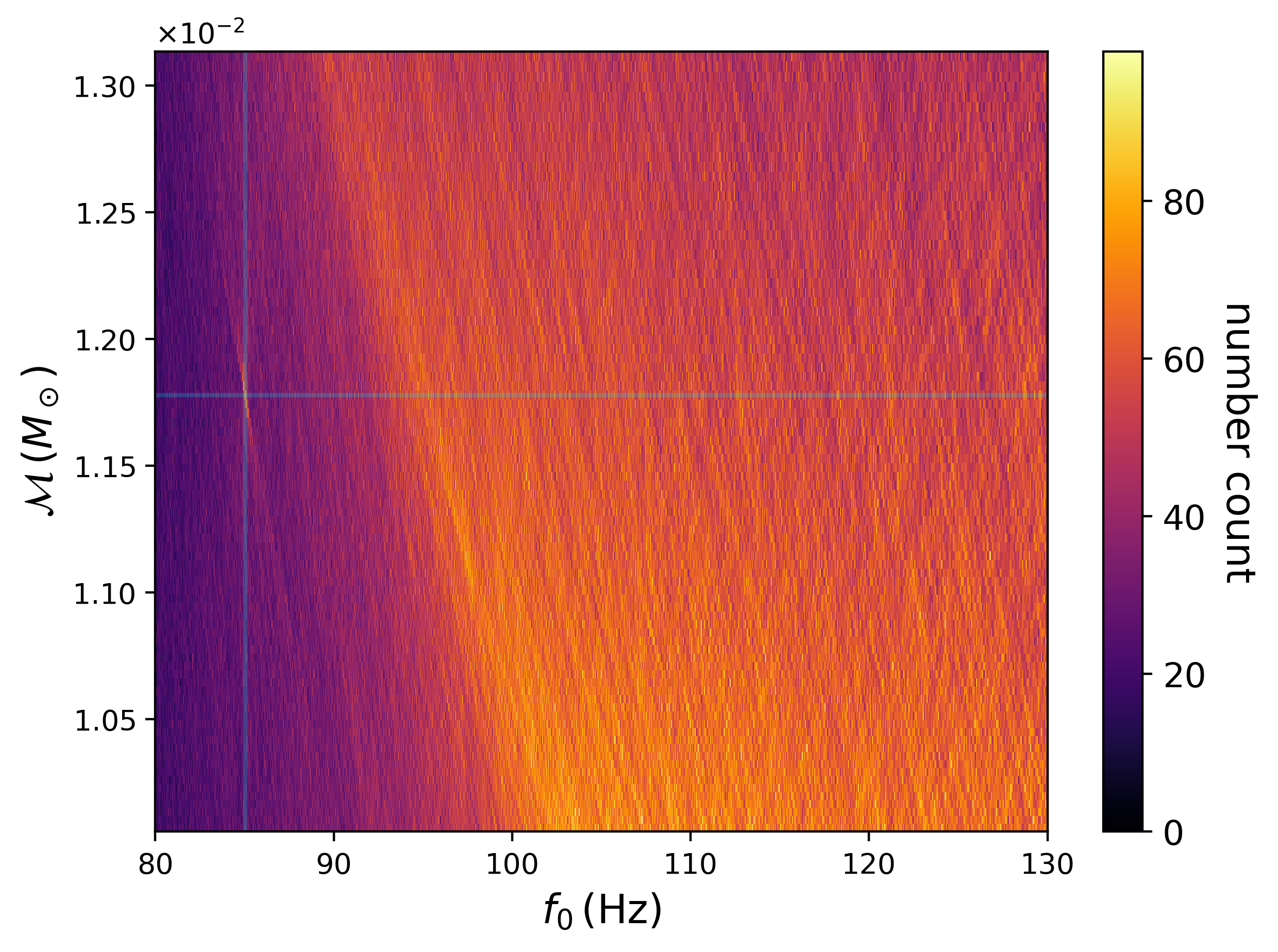}
%
    \end{center}
    \caption{\textbf{Limitation 2: injected signal power is spread over many $x_0$ bins in the \hm}. The maximum possible number count of this \pmap, created over a duration $\Tpm$ with overlapping FFTs of length $\TFFT$, is: $N_{\rm max}=2(\frac{\Tpm}{\TFFT})=4633$, but $<100$ peaks accumulate in the pixels surrounding the injection parameters. The intersection of the two blue lines indicate the pixel in which the injection should lie.
    }%
      \label{fig:hm-uni-inj-gauss-problem}
\end{figure}

We aim to address these two limitations in the following subsection by (1) computing the mean and standard deviation of the number of peaks expected in each pixel of the \hm, and (2) evaluating \cref{eqn:dx0} at each particular $\fgw$ analyzed, which creates a non-uniform grid in $x_0$ whose size matches the number of frequencies in the \pmap. We call our new method \BinGFH, and provide Matlab source codes in \cite{miller_2026_20599314}, and Python translations of them in \cite{andrew_l_miller_2026_20599034}. 

In \cref{subsec:expmean}, we explain the procedure to calculate the $CR$ in such a way as to guarantee that it follows a normal distribution in Gaussian noise. Then, in \cref{subsec:inj}, we show how this statistic behaves in the presence of a strong injection in Gaussian noise. Finally, in \cref{subsec:compcost}, we discuss the computational cost of the \BinGFH. 

\subsubsection{Expected mean and variance of Hough-map number counts in Gaussian noise}\label{subsec:expmean}

To model the statistical properties of the Hough number-count map in Gaussian noise, we compute the expected mean and variance of each pixel in the time--frequency plane. In each \hm, we evaluate the noise-only expectation value $\mu(x_0, k)$ and standard deviation $\sigma(x_0, k)$ for a set of templates defined on a generally nonuniform grid in the $(x_0, k)$ space.

For a given coherence time $\TFFT$, we construct the expected contribution from all FFTs and all frequency bins in the band $[\fmin, \fmax]$. The mean and variance of the Hough number count at each $(x_0, k)$ are obtained by summing over the discrete set of $(t, f)$ peaks that fall along each line in the \pmap. 

The probability $p_0$ of a peak surviving the thresholding procedure is empirically determined from the \pmap itself, as the ratio of the number of selected peaks to the total number of available frequency bins $N_{\rm bin}=(\fmax-\fmin)/\delta f$. This empirical value replaces the analytic estimate for $p_0$ (see \cref{sec:sens}) and ties the calculation of $\mu(x_0, k)$ and $\sigma(x_0, k)$ directly to the data. For each $(x_0, k)$, the expected mean and variance of the number count are then given by
\begin{align}
\mu(x_0, k) &= N_{\mathrm{pairs}}(x_0, k)\,p_0, \\
\sigma^2(x_0, k) &= N_{\mathrm{pairs}}(x_0, k)\,p_0(1 - p_0),
\end{align}
where $N_{\mathrm{pairs}}(x_0, k)$ is the number of time--frequency bins that project into the corresponding Hough bin.

In practice, we loop over all $k$ values and compute the mapping between $(t, f)$ and $(x_0,k)$ to determine $N_{\mathrm{pairs}}(x_0, k)$ per bin. The resulting matrices $\mu(x_0, k)$ and $\sigma(x_0, k)$ form the expected mean and standard deviation maps of the Hough transform, which are subsequently used, with the observed number counts $n(x_0, k)$, to compute the critical ratio:
\begin{equation}
CR(x_0, k) = \frac{n(x_0, k) - \mu(x_0, k)}{\sigma(x_0, k)}.
\end{equation}
In \cref{fig:hm-mean-gaus-noise}, we show the application of this procedure to a \hm created in Gaussian noise. From \cref{fig:hm_nonunix_gauss}, we can see that the the distribution of number counts is highly non-uniform. But, we can also compute the expected number count per pixel following the procedure outlined above, which is shown in \cref{fig:mu_nonunix_gauss}. By subtracting out this mean and dividing by an equivalent two-dimensional map of the standard deviation per pixel, we can obtain $CR(x_0,k)$ that has a mean of zero and a standard deviation of one, as shown in \cref{fig:CR-hist-gauss}. These \hms were created from \pmaps with a duration of 9266 s and $\TFFT=4$ s; however, we note that as the signal durations and $\TFFT$ we search for become shorter, the decrease in the number of peaks in the \pmap implies that the histogram in \cref{fig:CR-hist-gauss} becomes less normally distributed, as expected by the (opposite of) the central limit theorem. 

\begin{figure*}[ht!]
     \begin{center}
        \subfigure[ ]{%
            \label{fig:hm_nonunix_gauss}
            \includegraphics[width=0.49\textwidth]{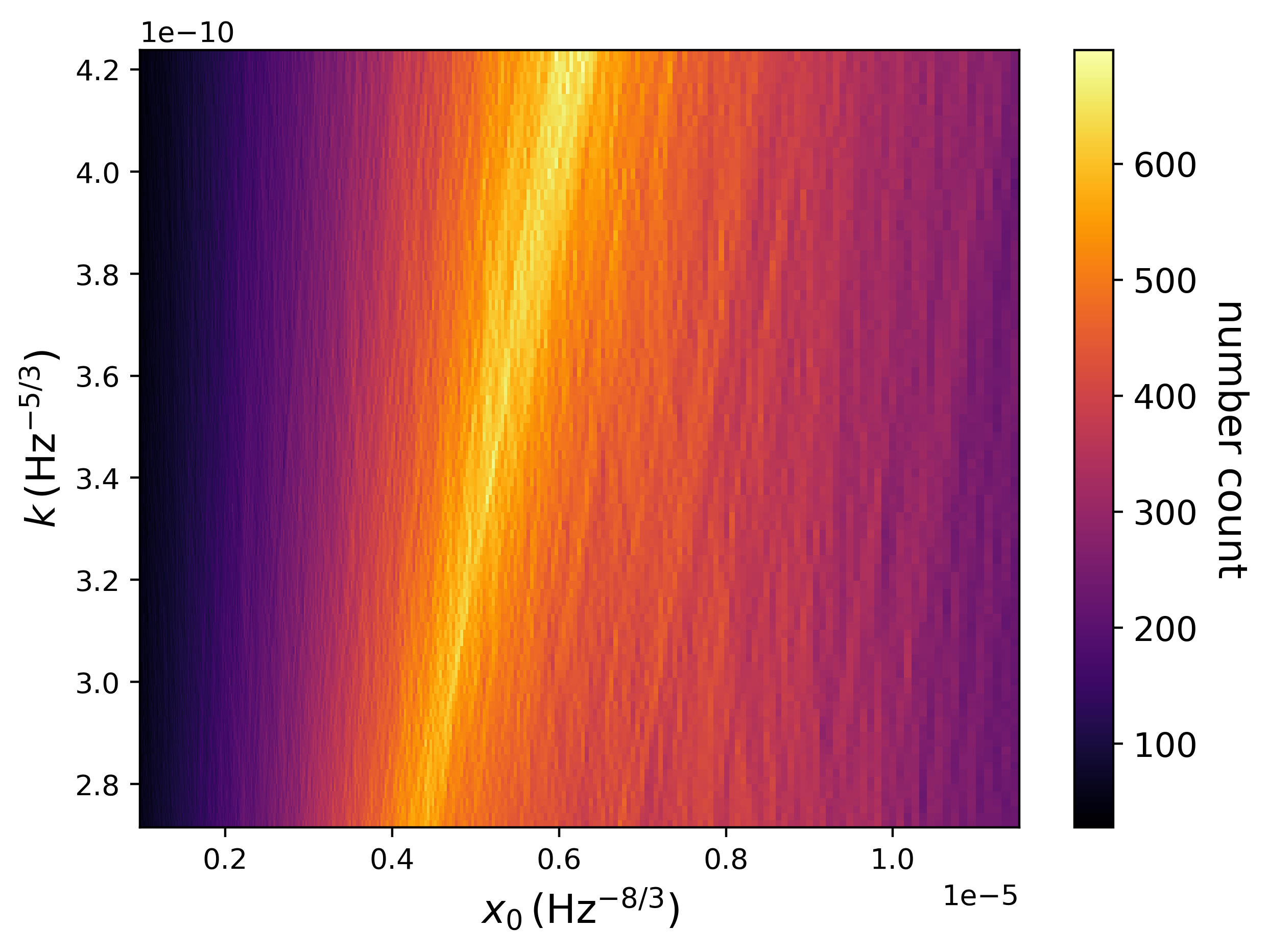}
        }%
        \subfigure[]{%
           \label{fig:mu_nonunix_gauss}
           \includegraphics[width=0.5\textwidth]{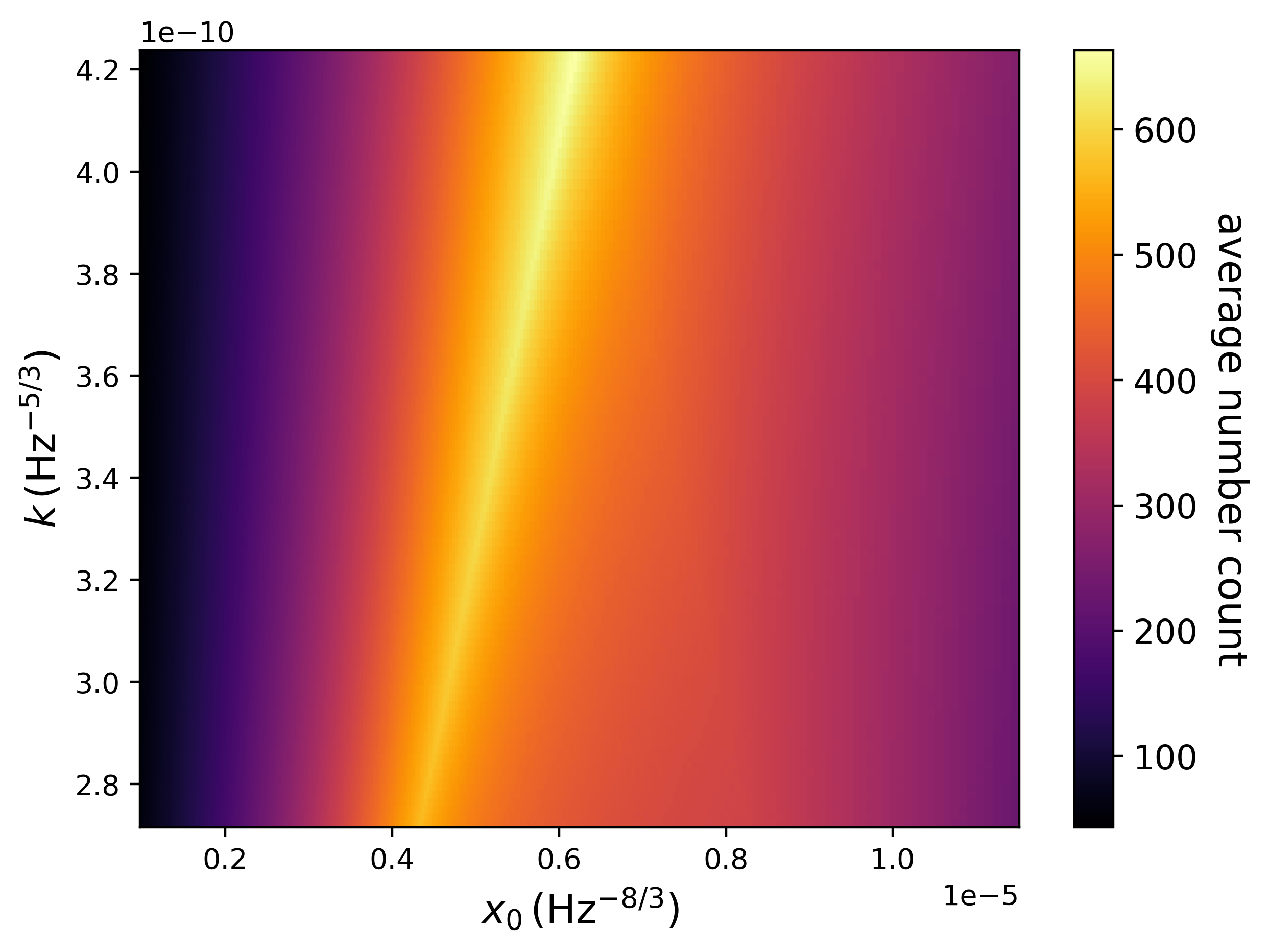}
        }\\ 
    \end{center}
    \caption[]{\textbf{\hm in Gaussian noise (left) and the estimation of the mean number count per pixel (right).} $\TFFT=4$ s; $\Tpm=9266$ s; $\Mc=[1,1.3]\times 10^{-2}\msun$; $f=[71,169]$ Hz. 
    }%
   \label{fig:hm-mean-gaus-noise}
\end{figure*}

\begin{figure}[ht!]
     \begin{center}
\includegraphics[width=0.49\textwidth]{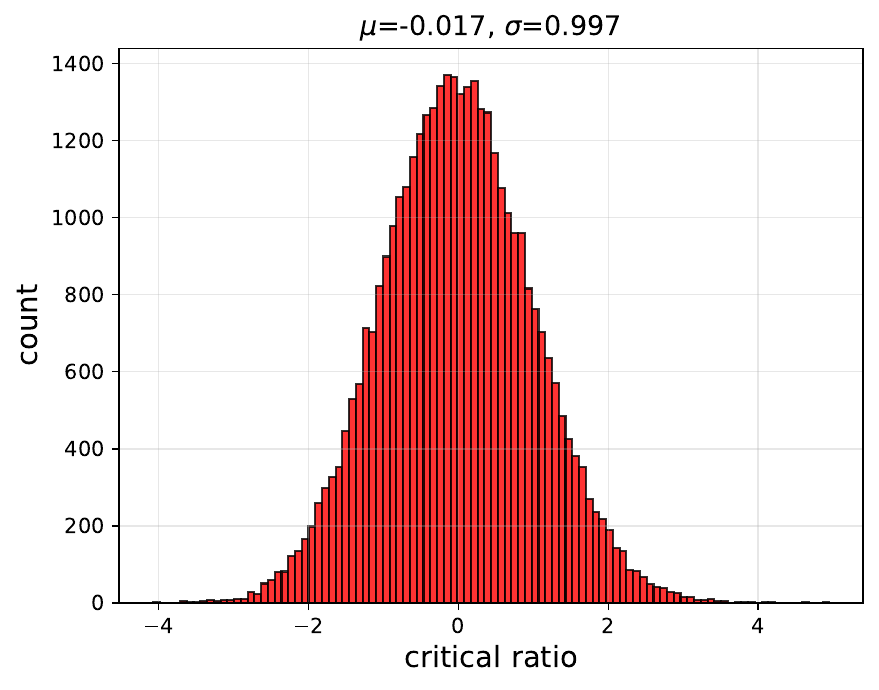}
%
    \end{center}
    \caption{\textbf{Critical ratio histogram obtained from Gaussian noise}. The critical ratio has been calculated from the \hm (\cref{fig:hm_nonunix_gauss}) by subtracting out the per-pixel mean (\cref{fig:mu_nonunix_gauss}) and dividing by the per-pixel standard deviation in the \hm.
    }%
      \label{fig:CR-hist-gauss}
\end{figure}


This procedure provides a data-driven, per-template noise normalization that naturally accounts for variations in duty cycle, frequency coverage, and the empirically measured peak probability $p_0(t)$. It also ensures that the $CR$ remains a random variable with a mean of zero and standard deviation of one. Ensuring that the $CR$ has these statistical properties implies that the candidates returned by the \BinGFH can be interpreted in a statistically meaningful way, paving the way for future detections with this method.

\subsubsection{$CR(x_0,k)$ in the presence of an injection in Gaussian noise}\label{subsec:inj}

It is important to understand how the statistical estimation behaves in the presence of a signal.  Given that the signal can contribute, at most, one peak per $\TFFT$, we expect that the normal distribution of the $CR$ should not change. However, the pixels in which the injection appear will have values of the $CR$ that are in the right-most tail of the distribution. We show an example of the \hm computed from a \pmap with an injection in Gaussian noise in \cref{fig:CR-hm-hist}. We have zoomed in on the signal parameters in order to show how well the \BinGFH recovers the injection in \cref{fig:hm_nonunix_gauss_inj}. Additionally, we show in \cref{fig:CR-hist-inj-gauss} the distribution of the $CR$, which shows the bulk normal distribution and a tail whose values correspond to the pixels around the parameters of the injection.

\begin{figure*}[ht!]
     \begin{center}
        \subfigure[ ]{%
            \label{fig:hm_nonunix_gauss_inj}
            \includegraphics[width=0.49\textwidth]{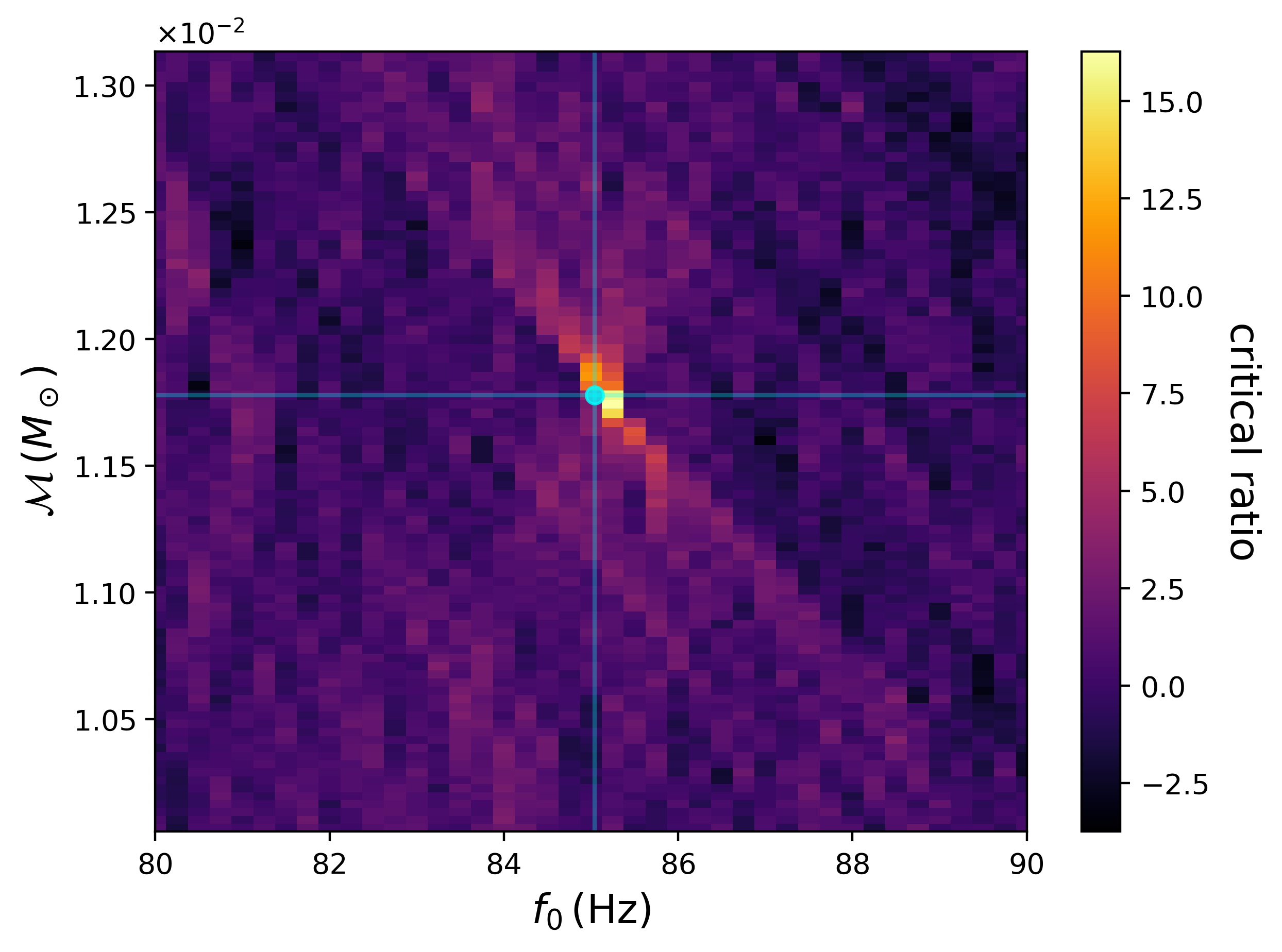}
        }%
        \subfigure[]{%
           \label{fig:CR-hist-inj-gauss}
           \includegraphics[width=0.5\textwidth]{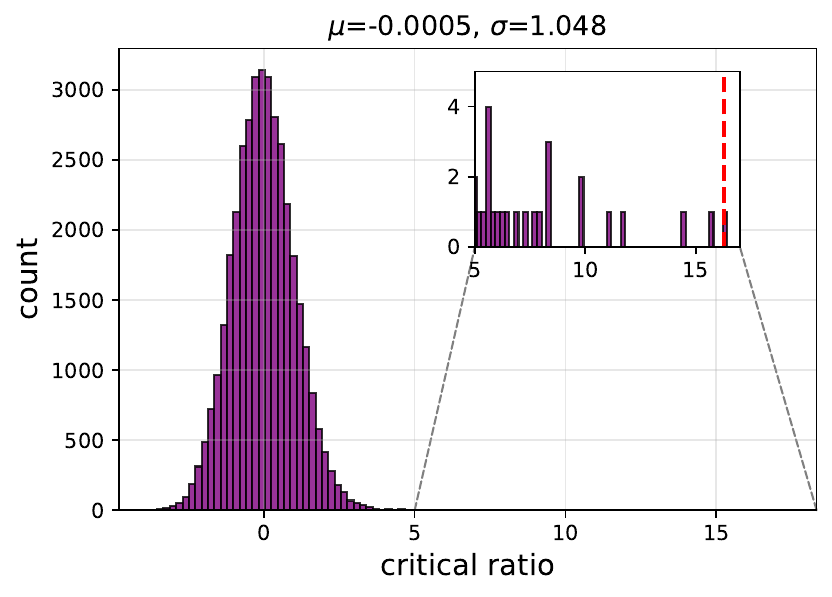}
        }\\ 
    \end{center}
    \caption[]{\textbf{\hm of an injection in Gaussian noise (left) and the histogram of the $CR$ values calculated from the \hm (right).} The $CR$ is calculated per pixel in the \hm by subtracting the number count by the per-pixel mean and dividing by the per-pixel standard deviation. The blue dot shows the injection parameters (left), and the inset shows a zoom of the number of critical ratios in the \hm between [5,15], all of which are caused by the injection (right). We note that, as a point of comparison, the maximum $CR$ in the \hm created by a uniform grid in $x_0$ is only $\sim 8$, because power is split among different across the too-finely-resolved grid in $x_0$
    }%
   \label{fig:CR-hm-hist}
\end{figure*}

\subsubsection{Computational cost}\label{subsec:compcost}

The \BinGFH leverages the same MATLAB functions that are used in the \GFHvtwo, permitting an order-of-magnitude speed-up with respect to the \GFHvone. However, the non-uniform grid in $x_0$ requires the additional MATLAB function \texttt{discretize}, which checks, for each point computed through \cref{eqn:dx0}, which bin it falls in. While very efficient, the use of \texttt{discretize} slows down \BinGFH by about 50\% with respect to \GFHvtwo for the $\TFFT=4$ s, $\Tpm=9266$ s configuration. However, for the shorter $\TFFT=[0.5,1]$ s, the computation time for both \GFHvtwo and \BinGFH is the same.

\section{Search design}\label{sec:search-design}

Following \cite{Alestas:2024ubs}, we compute the frequency range to analyze that optimizes the sensitivity towards inspiraling systems. This range is independent of the chirp mass or other signal parameters, and is computed per run by maximizing the following equation

\begin{equation}
    F(\fmin, \fmax) = \frac{\fmin^{2/3}}{\fmax^{11/24}} \sqrt{\int_{\fmin}^{\fmax} \frac{df}{f^{7/3} S_n (f)}}
    \label{eq:F_f0fs}
\end{equation}
over different values of starting frequency $\fmin$ and ending frequency $\fmax$. \cref{eq:F_f0fs} comes from plugging in the expected frequency dependence of the \gwh signal into the standard expression for \mfh \snr. Thus, maximizing \cref{eq:F_f0fs} also maximizes the \snr. 

For this study, we compute $\fmin$ and $\fmax$ for the \OFourA \psd, and show $F(\fmin,\fmax)$ in \cref{fig:f0-fstar}. For the O4a observing run, we find the optimal frequency range to analyze is $[\fmin,\fmax]=[71,169]$ Hz.

\begin{figure}[ht!]
     \begin{center}
\includegraphics[width=0.49\textwidth]{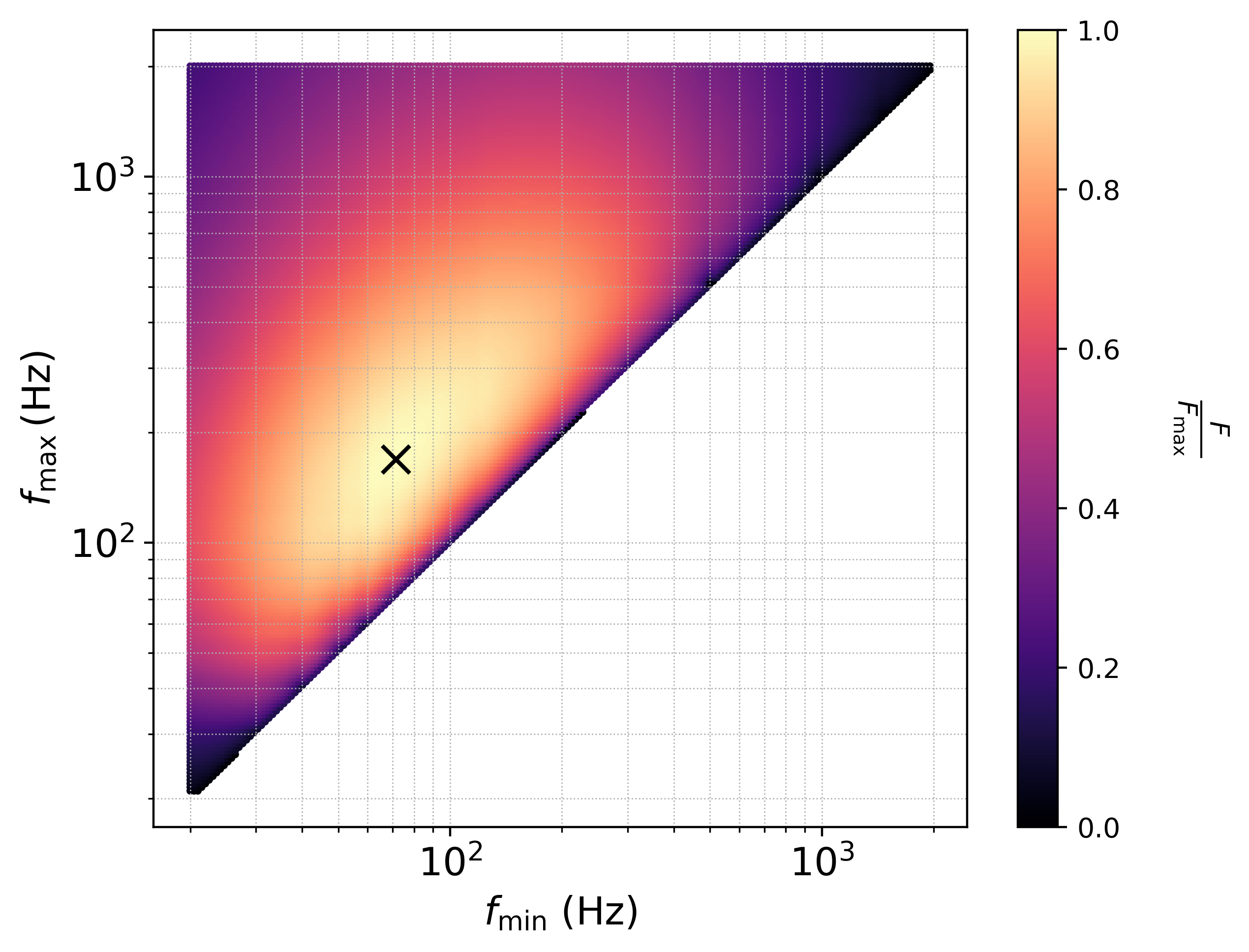}
%
    \end{center}
    \caption{\textbf{\cref{eq:F_f0fs} plotted for different choices of $\fmin$ and $\fmax$, showing a maximum at $[71,169]$ Hz}.
    }%
      \label{fig:f0-fstar}
\end{figure}

Noting that we wish to search over the chirp mass range $[10^{-2},10^{-1}]\msun$, we can then compute the corresponding minimum and maximum $\TFFT$ using \cite{Alestas:2024ubs}

\begin{equation}
    \TFFT^\mathrm{opt} = 8.50 ~ \mathrm{s} \left(\frac{\Mc}{10^{-2} \msun}\right)^{-5/6} \left(\frac{\fmax}{126.8 ~\mathrm{Hz}}\right)^{-11/6},
    \label{eq:TSFTopt}    
\end{equation}
in which we obtain $\TFFT^{\max}\simeq 5$ s and $\TFFT^{\min}\simeq 0.74$ s. Practically, we divide the range of chirp masses into four configurations, with $\TFFT=[4,3,2,1,0.5]$ seconds, and compute the range of chirp masses to which each configuration is optimally sensitive. The duration of each configuration is set by the maximum mass that each configuration can probe. These configurations are given in \cref{tab:configs}.

\begin{table}[h!]
\centering
\caption{Search configurations in the frequency band $[\fmin,\fmax]=[71,169]$ Hz.}
\begin{tabular}{cccc}
\hline
\hline
$\Tpm$ (s) & $\TFFT$ (s) & $\Mc^{\min}$ ($10^{-2} M_\odot$) & $\Mc^{\max}$ ($10^{-2} M_\odot$) \\
\hline
9226 & 4.0 & 1.00 & 1.31 \\
5190 & 3.0 & 1.31 & 1.85 \\
2307 & 2.0 & 1.85 & 3.02 \\
577  & 1.0 & 3.02 & 6.93 \\
314  & 0.5 & 6.93 & 10.0 \\
\hline
\hline
\end{tabular}
\label{tab:configs}
\end{table}

\section{Sensitivity}\label{sec:sens}

We would like to understand how the sensitivity of \BinGFH compares both to the sensitivity of \GFHvtwo and the theoretical sensitivity. Previous works have already shown that the \GFHvone and \GFHvtwo agree with theoretical predictions \cite{Miller:2018rbg,longpmr,Miller:2024jpo,LIGOScientific:2025vwc}; however, we note that such short signal durations and $\TFFT$ have not been tested. 

To define the sensitivty, we calculate the expected distance reach for which a fraction of injections $\Gamma$ would be recovered in a repeated number of experiments above a chosen threshold on the critical ratio $\crthr$:

\begin{equation}
    \dmax = D{\left(\crthr-\sqrt{2}\erfc^{-1}(2\Gamma)\right)^{-1/2}},
    \label{eqn:dmax}
\end{equation}
where $D$ denotes the distance away we could detect a signal as a function of the chirp mass, the frequencies covered by the signal and our analysis parameters $\TFFT,\Tpm$:

\begin{align}
    D&=1.41\left(\frac{G \mathcal{M}}{c^2}\right)^{5/3}\left(\frac{\pi}{c}\right)^{2/3} \frac{T_{\rm FFT}}{\sqrt{T_{\rm PM}}}  \nn\\ &\times\left(\sum_x^N \frac{f^{4/3}_{\text{GW},x}}{S_n(f_{\text{GW},x})}\right)^{1/2}  \left(\frac{p_0(1-p_0)}{Np^2_1}\right)^{-1/4}.\label{eqn:D}
\end{align}
$S_n$ is the noise \psd of \ifos, $N=\Tpm/\TFFT$, $\thetathr=2.5$ is the threshold for peak selection in the \pmap, and $p_0$ and $p_1$ are the probabilities of selecting a peak in the peakmap in the presence of Gaussian noise and a weak monochromatic signal, respectively, and can be calculated using $\thetathr$:

\begin{align}
    p_0&=e^{-\thetathr}-e^{-2\thetathr}+\frac{1}{3}e^{-3\thetathr},
    \label{eqn:p0} \\ 
        p_1 &= \thetathr \left( \frac{1}{2} e^{-\thetathr} - \frac{1}{2} e^{-2\thetathr} + \frac{1}{6} e^{-3\thetathr} \right) \nn \\ &+ \frac{1}{4} e^{-2\thetathr} - \frac{1}{9} e^{-3\thetathr}.
    \label{eqn:p1}
\end{align}
The minimum detectable strain amplitude at a chosen $\Gamma$, $\homin$, is simply:

\begin{equation}
    \homin = \frac{4}{\dmax} \left( \frac{G \Mc}{c^2} \right)^{5/3}
\left( \frac{\pi \fgw}{c} \right)^{2/3},
\label{eqn:h0min}
\end{equation}
where $\fgw$ is taken to be the maximum frequency of the signal within $\Tpm$. 

\tcb{The above equations are derived assuming Gaussian, stationary noise. If there is significant deviation from Gaussianity across multiple frequencies, then injections must be performed to assess the sensitivity of our method and be compared with \cref{eqn:h0min}, as was done in \cite{LIGOScientific:2025vwc}.}

\subsection{Comparison}

We perform 50 injections at 35 amplitudes ranging from $[10^{-24},10^{-22}]$ in Gaussian noise with a constant \psd of $\sqrt{S_n}=7.94\times 10^{-24}\hz^{-1/2}$ to determine how the efficiency of the \BinGFH changes as a function of signal strength. One chirp mass per configuration in \cref{tab:configs} is randomly selected, and nuisance parameters (polarization angle, $\cos\iota$, and sky position) are randomized for each of the 50 injections. 

The efficiency curves for each of the configurations are shown in \cref{fig:eff-curves}. We observe a sigmoid shape for the efficiency, consistent with what is expected in Gaussian noise. The sensitivity gets slightly worse in strain for shorter $\TFFT$ and $\Tpm$, despite the fact that $\Mc$ is increasing. However, because the distance reach depends strongly on $\Mc$, we are able to see farther away for heavier systems, as expected. This will be discussed further in the next subsection.

Additionally, we show in \cref{fig:h0-d95-comparison} a comparison of the minimum detectable strain amplitude and distance reach at 95\% confidence as a function of chirp mass, for \GFHvtwo, \BinGFH and the predicted value from \cref{eqn:h0min,eqn:dmax}. We see generally good agreement between the three scenarios, but note that \GFHvtwo cannot achieve 95\% detection efficiency for systems with the largest chirp mass. Moreover, we note that the error bars on the theoretical points correspond to correcting \cref{eqn:h0min,eqn:dmax} for the source-specific parameters used in the injection \cite{KAGRA:2022dwb}: note that \cref{eqn:dmax,eqn:h0min} are derived for population-averaged parameters (in $\cos\iota$, polarization angle and sky position). Thus, we multiplied (divided) \cref{eqn:h0min} (\cref{eqn:dmax}) by the mean of the ratio between the source-specific amplitude and the population-averaged amplitude, and the error bars on the theoretical points represent $\pm 1\sigma$. See the appendices of \cite{LIGOScientific:2025vwc} for further details on this source-specific factor. In contrast, the error bars on the empirical and \GFHvtwo correspond to the binomial uncertainty on the 95\% efficiency.

\begin{figure}[ht!]
     \begin{center}
\includegraphics[width=0.49\textwidth]{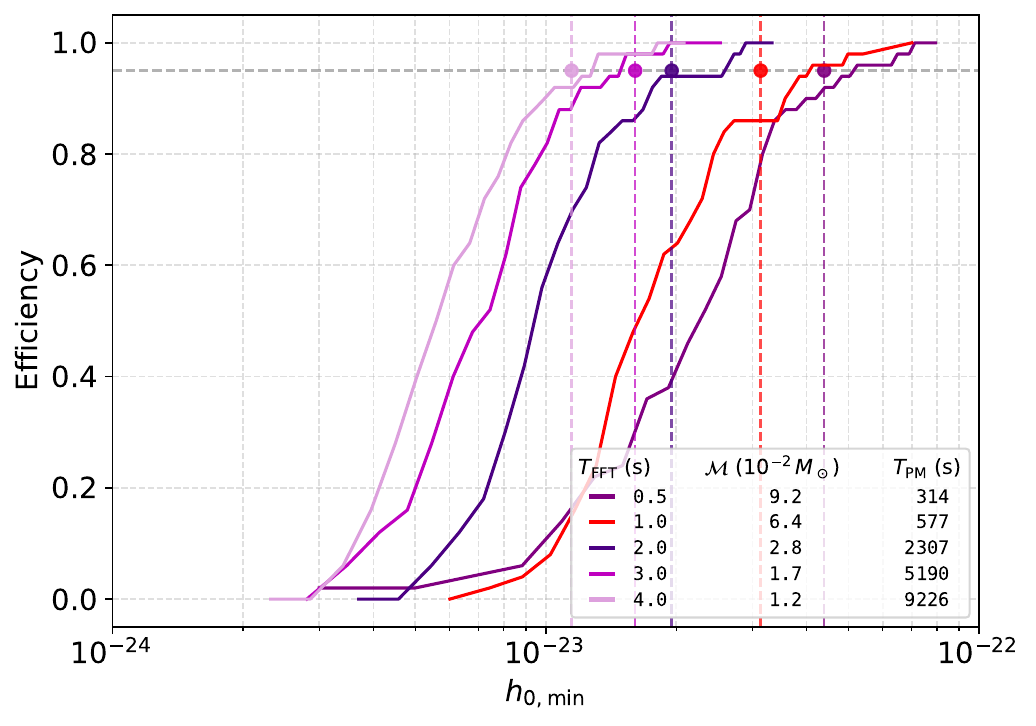}
%
    \end{center}
    \caption{\textbf{Efficiency curves for \BinGFH for different chirp masses, coherence times and signal durations}. Fifty injections per amplitude were performed in Gaussian noise at a level of $\gnlvl$, randomized over nuisance parameters (sky position, cosine of the inclination angle, and polarization). Individual points indicate the 95\% confidence-level $h_{\rm 0,min}$ for each configuration.
    }%
      \label{fig:eff-curves}
\end{figure}

\begin{figure*}[ht!]
     \begin{center}
        \subfigure[]{%
            \label{fig:h095}
            \includegraphics[width=0.49\textwidth]{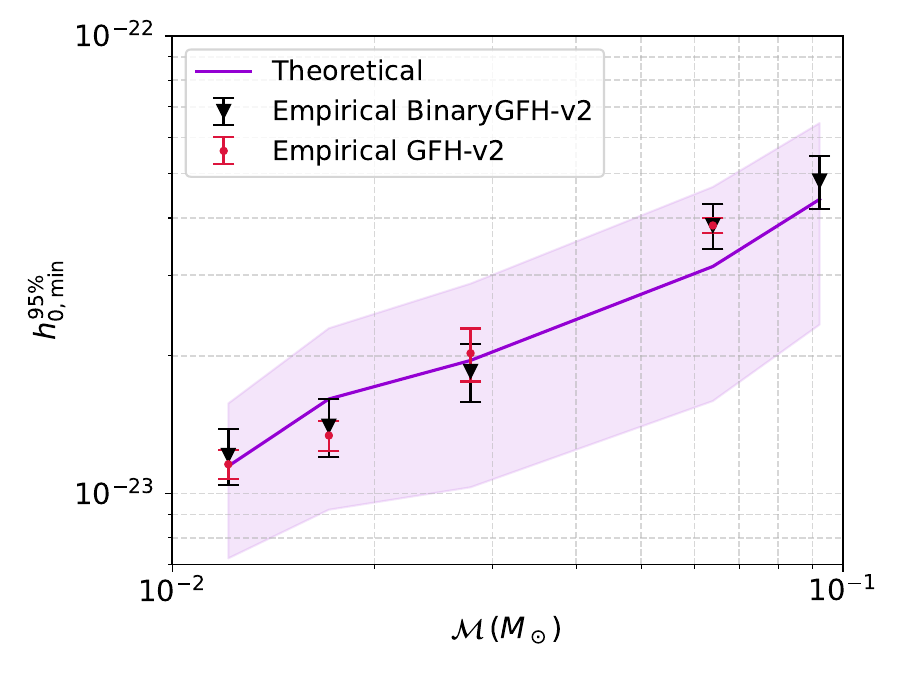}
        }%
        \subfigure[]{%
           \label{fig:d95}
           \includegraphics[width=0.48\textwidth]{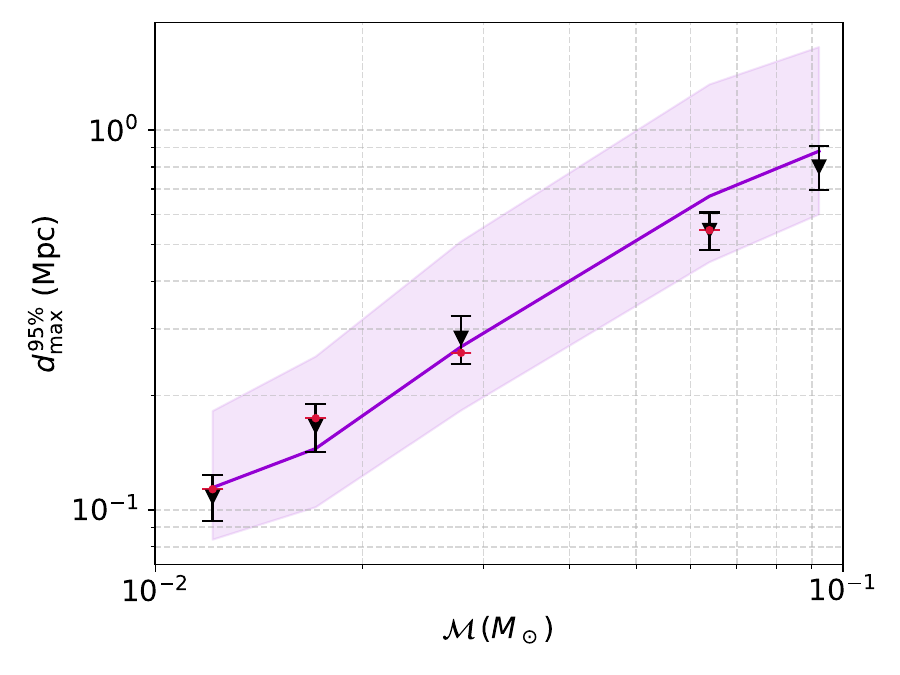}
        }\\ 
    \end{center}
    \caption[]{\textbf{Comparison at 95\% confidence level of the minimum detectable amplitude (left) and maximum distance reach (right) of \GFHvtwo (red) and \BinGFH (black) versions of the \gfh, and the theoretical expectation (purple).} 
    The injections used here are the same as those from \cref{fig:eff-curves}. 
    We note that \GFHvtwo did not achieve 95\% efficiency for the largest chirp mass (and shortest $\TFFT=0.5$\,s and $\Tpm=314$\,s) presented here. 
    The error bars on the “Empirical” curves represent the binomial error, while those on the “Theoretical” curve correspond to a correction of \cref{eqn:dmax} for the source-specific parameters used in obtaining the empirical sensitivity estimate.}%
   \label{fig:h0-d95-comparison}
\end{figure*}

\subsection{Space-time volume and rate density}

With the efficiency curves, we can compute the space-time volume $\avgVT$ to which a search would be sensitive. For a fixed chirp mass and frequency range, the only parameter that affects the efficiency of the search is the \gwh amplitude (or distance) from the \ifos, so $\avgVT$ can be written as

\begin{equation}
    \avgVT = \Tobs \int_{0}^{\infty} 4 \pi r^{2} \, \epsilon(r) \, dr ,
    \label{eqn:intvt-app}
\end{equation}
where $\epsilon(r)$ is the efficiency as a function of an arbitrary distance $r$. The space-time volume is thus the integral over the efficiency functions given empirically in \cref{fig:eff-curves}, which essentially down-weight the distances as they become larger and larger.

It was noted in \cite{LIGOScientific:2025vwc} that, assuming that the $CR$ follows a normal distribution, the efficiency function can be approximated as:

\begin{equation}
    \epsilon(r) = P(CR > \crthr \mid r) = \frac{1}{2} \, \text{erfc} \left( \frac{\crthr - \left(\frac{D}{r}\right)^2}{\sqrt{2}} \right)
    \label{eqn:eff_D}
\end{equation}
Then, the co-moving space-time volume $\avgVT$ in \cref{eqn:intvt-app} can be found by using Laplace's method, which considers only the asymptotic contributions of the integral:

\begin{align}
     \avgVT 
     \simeq\Tobs\frac{4}{3}\pi \left(\frac{D}{\sqrt{\crthr}}\right)^3
     \label{eqn:avgvt_laplace}
\end{align}
Assuming that the event rate for inspiraling ultra-compact objects is Poissonian, consistent with \ssm searches \cite{Phukon:2021cus,LIGOScientific:2021job,LIGOScientific:2022hai}, we can then calculate the upper limits on the rate density at a chosen confidence level $\alpha=0.9$:
\begin{equation}
    \mathcal{R}_{\rm 90\%} = \frac{2.303}{\avgVT}.\label{eqn:R90}
\end{equation}
We compare these two approaches for calculating $\avgVT$ and the corresponding rate densities in the parameter space $[10^{-2},10^{-1}]\,\msun$ in \cref{fig:VT-R-comparison}. The two approaches are: (1) integrating over the efficiency curve (``Full'') and (2) approximating the integral using Laplace's method (``Laplace''). We note that the approximation gives slightly worse, i.e. more conservative results, in Gaussian noise, which indicates its applicability as a way of avoiding extensive injection campaigns in this parameter space, as well as in wider ones such as those analyzed in \cite{Miller:2024fpo,LIGOScientific:2025vwc}.

\begin{figure*}[ht!]
     \begin{center}
        \subfigure[ ]{%
            \label{fig:VT-comparison}
            \includegraphics[width=0.49\textwidth]{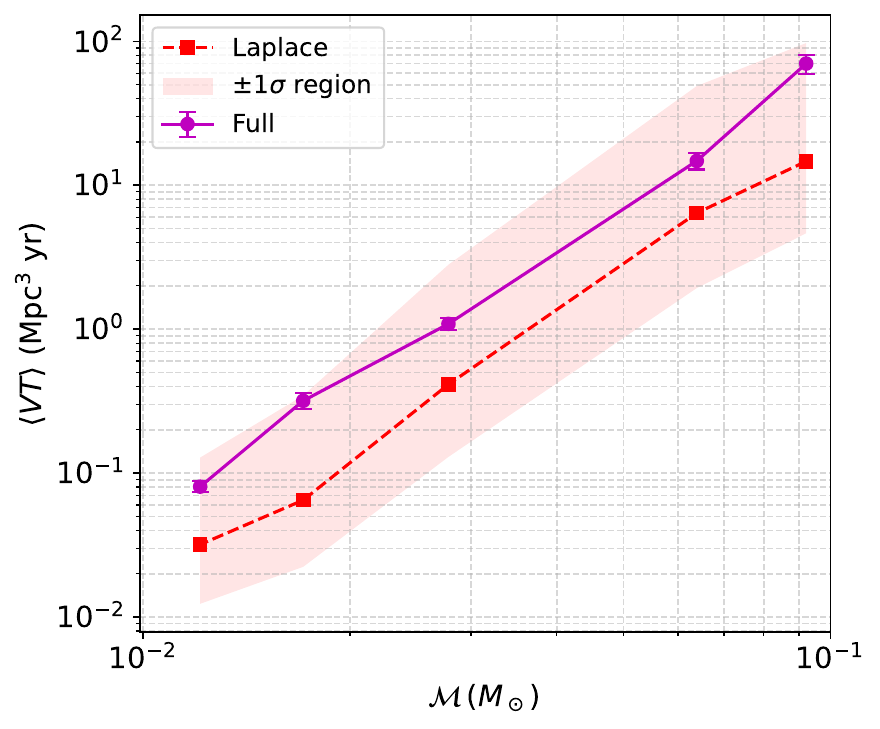}
        }%
        \subfigure[]{%
           \label{fig:R90-comparison}
           \includegraphics[width=0.49\textwidth]{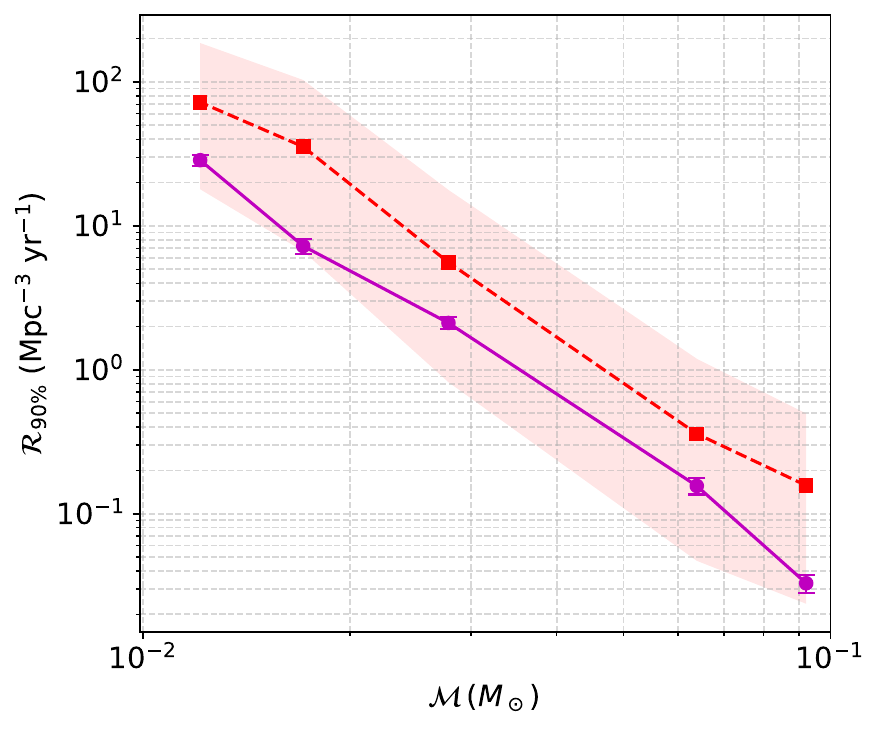}
        }\\ 
    \end{center}
    \caption[]{\textbf{Comparison of the calculation of $\avgVT$ and $\mathcal{R}_{\rm 90\%}$ using the full efficiency curve and Laplace's method.}
    }%
   \label{fig:VT-R-comparison}
\end{figure*}

\section{Projected constraints}\label{sec:proj-constr}

{Placing limits on the \pbh abundance is difficult and highly model-dependent.  Indeed, several binary formation channels have been proposed and each of them depend on the \pbh mass distribution and are subject to multiple astrophysical uncertainties.  

Following state-of-the art rate prescriptions~\cite{Raidal:2024bmm}, early-universe two-body binaries are typically the dominant binary formation channel for the masses and \dm fraction relevant for this work.  Their merger rate densities $\mathcal{R}^{\rm cos}_{\rm prim}$ are given by }  
\begin{align}
        \mathcal{R}^{\rm cos}_{\rm prim}  &\approx   {1.6 \times 10^{-12}}{\text{ \invkpccubedyr}} \ftilde^{53/37} \nonumber \\  &\times \left(\frac{m_1 + m_2}{\,\msun}\right)^{-32/37}\left[\frac{m_1 m_2}{(m_1+m_2)^2}\right]^{-34/37},  \label{eq:cosmomerg}
\end{align}
where we define an effective parameter $\ftilde$ as:

\begin{equation}
\ftilde \equiv \fpbh \left[\fsup\fmone\Delta \ln m_1\fmtwo\Delta \ln m_2\right]^{37/53},
\label{eqn:ftilde}
\end{equation}
This effective parameter encodes the major sources of uncertainty on \pbh constraints: the mass functions $\fmone$ and $\fmtwo$, the suppression factor $\fsup$, and the fraction of \dm that \pbhs could compose $\fpbh$.

{If the galactic \dmh density was the one at the Sun's location, $\rho_{\rm DM}\simeq10^{16} \,\msun {\rm Mpc}^{-3}$ \cite{Weber:2009pt}, the merger rates would be enhanced to $\mathcal{R} = 3.3 \times 10^5 \mathcal{R}_{\rm prim}^{\rm cos}$~\cite{Miller:2020kmv}.  However, with current-generation \gwh \ifos data, we can probe source distances comparable to the distance to the Galactic Center from Earth. We therefore have allowed for this enhancement of the rates but also decrease them proportionally by a factor $F(d)$ that accounts for the integrated \dmh density profile centered on the sun location at $8.2$ kpc from the galactic center, as described in \cite{LIGOScientific:2025vwc}. For future \gwh \ifos, however, no such enhancement factor is added because the distance reach of such searches will be at the level of tens to hundreds of megaparsecs.}

We now turn to project constraints on $\ftilde$ and $\fpbh$ using current-generation and next-generation \gwh \ifos. In particular, we use the \psds from LIGO O4a Livingston and \ce to calculate the expected distance reaches (\cref{eqn:D}) and rate densities (\cref{eqn:R90}) in these observing runs using $\crthr=7$ for each of the configurations listed in \cref{tab:configs}. For O4a, we use the actual observation time of $\Tobs=237$ days; for \ce, we set $\Tobs=4$ years, and keep the same $[\fmin,\fmax]=[71,169]$ Hz.

We show in \cref{fig:ftilde-equal-asymm} constraints on the \pbh model-agnostic parameter $\ftilde$ for both O4a and \ce for both equal-mass and asymmetric mass-ratio systems. We take $m_1=2.5\msun$ in the latter case, motivated by their formation during the QCD phase transition \cite{Carr:2019kxo}. We can see that, for equal-mass systems, $\ftilde<1$ is only reachable in \ce, while, for asymmetric mass-ratio systems, $\ftilde<1$ across all $m_2$ considered in both detectors considered. 

For particular choices of suppression factors and mass functions, we also present projected constraints on $\fpbh$ in \cref{fig:fpbh-equal-asymm}. We assume a monochromatic mass function in the equal-mass case, and equal proportions of $m_1$ and $m_2$ in the asymmetric mass-ratio case. Furthermore, we vary the suppression factor from no suppression ($\fsup=1$, best-case scenario) to severe suppression ($\fsup=2.3\times10^{-3}\fpbh^{-0.65}$, worst-case scenario) \cite{Hutsi:2020sol}. In current \gwh data, we can only constrain $\fpbh<1$ for minimal or negligible rate suppression; however, \ce will allow us to place physically relevant constraints on $\fpbh$ even in the worst-case scenario.

\begin{figure*}[ht!]
     \begin{center}         \includegraphics[width=\textwidth]{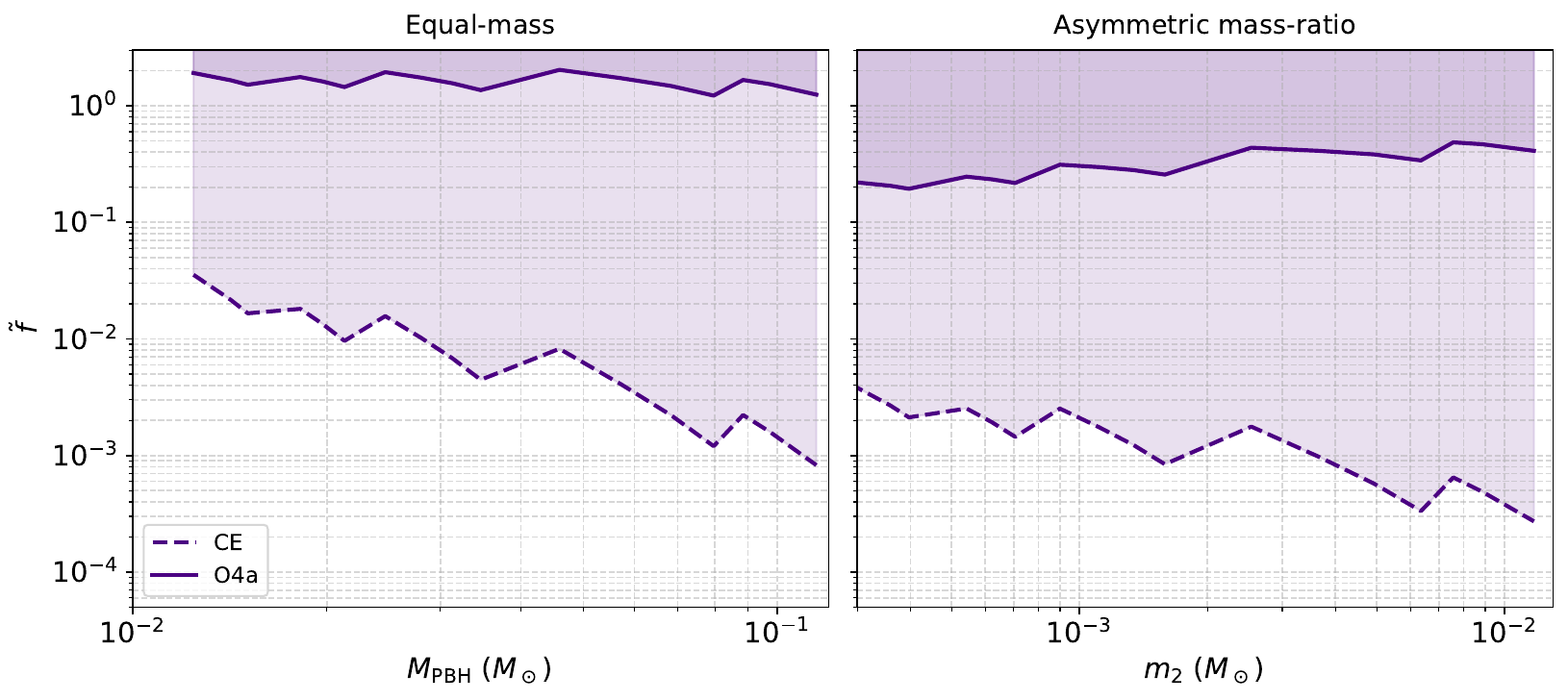}
    \end{center}
    \caption[]{\textbf{Projected constraints in the current observing run (O4a) and \ce on the model-agnostic parameter $\ftilde$ for equal-mass (left) and asymmetric mass-ratio (right) systems.} $\mpbh=2^{1/5}\Mc$. For O4a, $\Tobs=237$ days; for \ce, $\Tobs=4$ years.
    }%
   \label{fig:ftilde-equal-asymm}
\end{figure*}

\begin{figure*}[ht!]
     \begin{center}         \includegraphics[width=\textwidth]{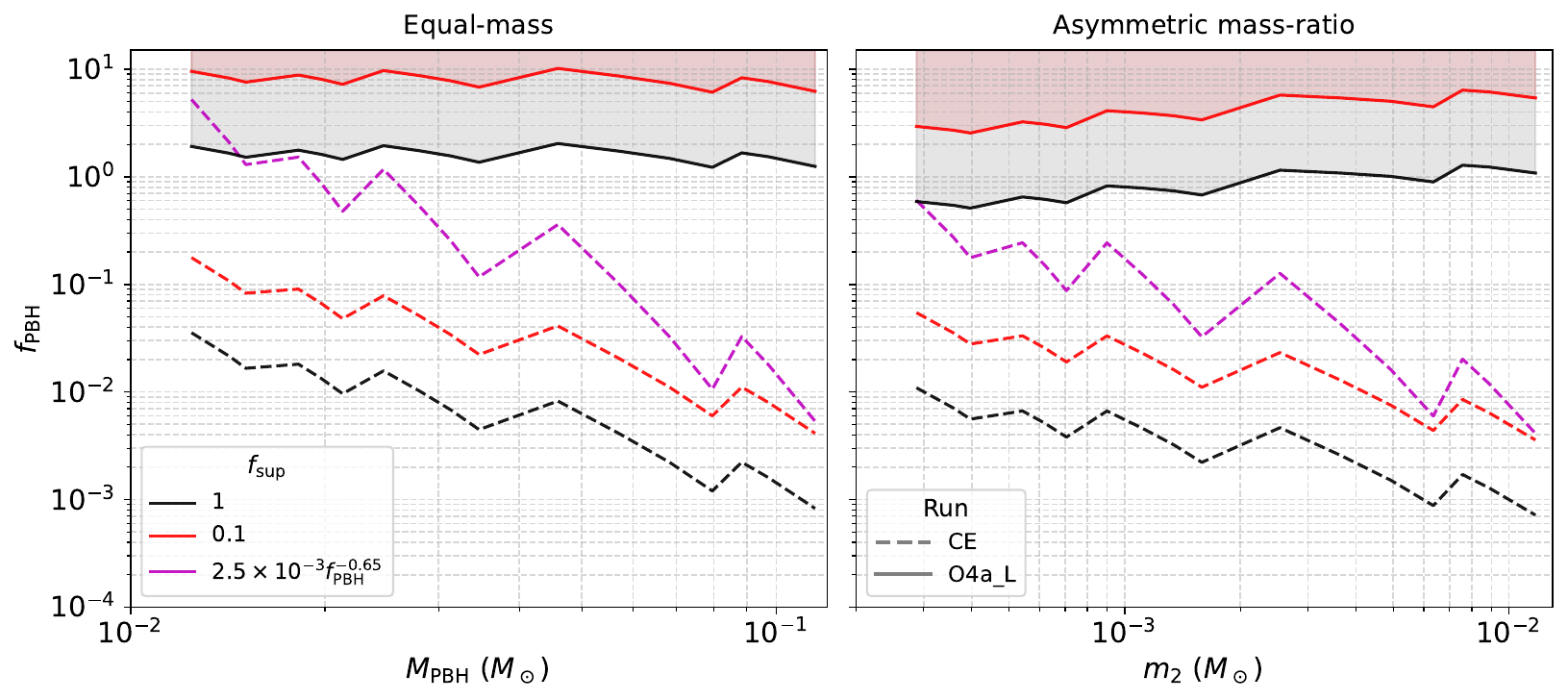}
    \end{center}
    \caption[]{\textbf{Projected constraints in the current observing run (O4a) and \ce on the $\fpbh$ for equal-mass (left) and asymmetric mass-ratio (right) systems.} $\mpbh=2^{1/5}\Mc$. For O4a, $\Tobs=237$ days; for \ce, $\Tobs=4$ years. For equal-mass systems, we take $\fmone = \fmtwo = 1$; for asymmetric mass-ratio systems, we take $\fmone=\fmtwo=0.5$ and $m_1=2.5\msun$. In both cases, we show how the constraints on $\fpbh$ change when we vary $\fsup$. $\fsup=1$ indicates no suppression of binary formation.
    }%
   \label{fig:fpbh-equal-asymm}
\end{figure*}

\section{Conclusions}\label{sec:concl}

We developed a new method, called \BinGFH, that can detect \gws from inspiraling ultra-compact objects whose chirp masses lie in the range $\Mc=[10^{-2},10^{-1}]\msun$. To do so, we have refined the \GFHvtwo to work with a non-uniform grid in $x_0$. We have also improved the statistical robustness of this method by estimating the expected background for the number counts in the \hm due to the nonlinear transformation of the \pmap (\cref{eqn:xx0}). Both of these improvements are not unique to the study presented here: in fact, any search that relies on the \gfh, e.g. searches for long-lived remnants of \bnsh mergers or supernovae, can apply these improvements, albeit for spinning down \nss with different $n=[3,5,7]$, depending on the dominant spin-down mechanism.

To validate our improved method, we perform injections at various chirp masses that would correspond to a realistic search that we could do in O4a data. We show that our new method, \BinGFH, agrees well with theoretical expectations -- both in expected distance reach and rate density --, and improves upon \GFHvtwo by being able to handle systems with $\fdot\lesssim4$ Hz/s (\GFHvtwo can only find systems with $\fdot\lesssim 1$ Hz/s). We project constraints that we would obtain on the fraction of \dm that \pbhs can compose with current-generation and future-generation \gwh detectors, both of which indicate that our method can be sensitive to a physical regime in which $\fpbh<1$ depending on the suppression factor.

\BinGFH bridges the gap between previous searches for \cws and \tcws for inspiraling ultra-compact objects with $\Mc=[10^{-7},10^{-2}]\msun$ and \mfh searches that target $\Mc=[10^{-1},1]\msun$. Though $\fpbh<1$ is constrained in this mass regime with microlensing experiments, it is essential to have multiple probes of \pbhs across the mass parameter space, because all of the constraints make different assumptions regarding the formation mechanisms of \pbhs. In particular, \gwh searches can probe the scenario in which \pbhs form in binaries, while microlensing analyses can handle the case in which \pbhs are sufficiently isolated from one another. Thus, our method and proposed search fits well within the paradigm of probing the existence of \pbhs or other ultra-compact objects in a variety of ways.

\input{acknowledgements}

\bibliographystyle{apsrev4-1}
\bibliography{references} 
\end{document}

%% file: acknowledgements.tex
\section*{Acknowledgments}
This material is based upon work supported by NSF's LIGO Laboratory which is a major facility fully funded by the National Science Foundation.

This work is supported by the National Natural Science Foundation of China (NSFC) under Grant No. 12347103 and 12547104.

We would like to thank the Rome Virgo group for the tools necessary to perform these studies, such as the development of the original \fh transform and the development of the short FFT databases. Additionally we would like to thank Luca Rei for managing data transfers.

This research has made use of data, software and/or web tools obtained from the Gravitational Wave Open Science Center (https://www.gw-openscience.org/ ), a service of LIGO Laboratory, the LIGO Scientific Collaboration and the Virgo Collaboration. LIGO Laboratory and Advanced LIGO are funded by the United States National Science Foundation (NSF) as well as the Science and Technology Facilities Council (STFC) of the United Kingdom, the Max-Planck-Society (MPS), and the State of Niedersachsen/Germany for support of the construction of Advanced LIGO and construction and operation of the GEO600 detector. Additional support for Advanced LIGO was provided by the Australian Research Council. Virgo is funded, through the European Gravitational Observatory (EGO), by the French Centre National de Recherche Scientifique (CNRS), the Italian Istituto Nazionale della Fisica Nucleare (INFN) and the Dutch Nikhef, with contributions by institutions from Belgium, Germany, Greece, Hungary, Ireland, Japan, Monaco, Poland, Portugal, Spain.

We also wish to acknowledge the support of the INFN-CNAF computing center for its help with the storage and transfer of the data used in this paper.

We would like to thank all of the essential workers who put their health at risk during the COVID-19 pandemic, without whom we would not have been able to complete this work.

This work is partially supported by ICSC – Centro Nazionale di Ricerca in High Performance Computing, Big Data and Quantum Computing, funded by European Union – NextGenerationEU.